\documentclass[12pt]{article}
\usepackage[cp1251]{inputenc}
\usepackage[T2A]{fontenc}
\usepackage[OT1,T2A]{fontenc}
\usepackage[english,russian]{babel}

\usepackage{amsmath,amssymb}
\usepackage{amsfonts,
amssymb,amsmath,
latexsym
}
\usepackage{mathrsfs}

\evensidemargin\oddsidemargin \topmargin2cm

\usepackage{amsfonts,amscd,amsmath}

\renewcommand{\baselinestretch}{1.32}
\usepackage{mathrsfs} 

\begin{document}
\large

\renewcommand{\baselinestretch}{1.32}
\sloppy



\begin{center}
─ц.├юЇ (J.E. Gough)$^1$, ╥. ╤.╨рЄ№■( T.S. Ratiu)$^{2}$,
╬.├.╤ьюы эют (O.G. Smolyanov)$^{3}$

{\bf   Wigner Measures and Quantum Control}
\end{center}
\addtocounter{footnote}{1} \footnotetext{Institute of Mathematics,
Physics and Computer Sciences, Aberystwyth University, Great
Britain \\ \texttt{jug@aber.ac.uk}}\addtocounter{footnote}{1} \footnotetext{Section de
Math\'ematiques and Bernoulli Center, \'Ecole Polytechnique
F\'ed\'erale de Lausanne. CH--1015 Lausanne. Switzerland.
\texttt{tudor.ratiu@epfl.ch} } \addtocounter{footnote}{1}
\footnotetext{╠хїрэшъю-ьрЄхьрЄшўхёъшщ Їръєы№ЄхЄ ╠юёъютёъюую
уюёєфрЁёЄтхээюую єэштхЁёшЄхЄр шь. ╠.┬. ╦юьюэюёютр.
\texttt{smolyanov@yandex.ru} }

\begin{abstract}
We study several examples from quantum control theory (see \cite{GoJa2009,GoJaMaMoWa2012}) in the framework of Wigner functions and measures for infinite dimensional open quantum systems. An axiomatic definition of coherent quantum feedback is proposed within this setting.
\end{abstract}

\date{}


\makeatother







The representation of the states of quantum systems
in terms of Wigner measures allows for useful analogies with classical system dynamics; 
it is similar to the representation of the states of classical Hamiltonian systems in
terms of probability measures on the phase space. The passage to a description of the state of
a subsystem of some larger quantum system is in both cases implemented by means of projective operations, because
the phase space of the classical analogue of the ambient quantum system, being the union of some subsystems, is the Cartesian product of the phase spaces of
the classical analogues of these subsystems.
In the more familar setting of a finite dimensional phase space we can consider,
instead of Wigner measures, their
densities with respect to the Liouville measure, which
are classical Wigner functions. However, on an infinite dimensional phase space, there exists no Liouville measure, i.e., a Borel  $\sigma$-additive  $\sigma$-finite locally
finite measure invariant with respect to symplectic
transformations (this is a consequence of a well known
theorem of Weil). In this case, we can either directly
apply the Wigner measure, or introduce some Уsufficiently niceФ measure instead of the Lebesgue measure. In the case of a linear phase space, we can
use a Gaussian measure as the Уsufficiently niceФ measure, as is done in the so-called
white noise analysis. After this, it becomes possible
again to replace Wigner measures by \lq\lq Wigner functions\rq\rq , i.e., by their densities with respect to the new
measure. We shall consider Wigner measures and their
densities in parallel.

Section 1, which is of independent interest,
considers the properties of Wigner measures and functions; some of the results of this section can be
regarded as an extension of results of [4] to Wigner
measures. In section 2, an equation describing
the evolution of the Wigner functions of quantum systems obtained by quantizing Hamiltonian systems
with infinite dimensional phase space is given; this
equation is obtained as a consequence of a similar
equation for the evolution of a Wigner measure
(see [5]). (A Wigner measure is a signed cylindrical
measure, and it would be interesting to estimate its
variation and find countable additivity conditions;
however, we do not discuss these issues here.) 

The final section considers the evolution of the Wigner
measures and functions of subsystems of quantum systems. Here, models of control of 
quantum systems are discussed and an axiomatic definition
of coherent quantum feedback is given, which, as far
as we know, has not been explicitly introduced in the
literature so far. We consider largely algebraic aspects
of the theory, omitting analytical assumptions.

{\bf 1. WIGNER MEASURES
AND FUNCTIONS.}

This section discusses properties of Wigner measures and their densities with respect to fixed measures
on a classical phase space, that is, Wigner functions
(precise definitions are given below). Let $ E :=  Q \times  P$ be
the phase space of a Hamiltonian system, where $ Q$ and $P$ are real locally convex spaces (LCSs),  $P =  Q^\ast$, and
$Q =  P^\ast$ (given an LCS  $X$ , we denote by $X^\ast$ its dual endowed with a locally convex topology consistent
with the duality between  $X$ and  $X^\ast$ ); then  $E^\ast =  P \times  Q$.
Suppose also that  $\langle  \cdot , \cdot \rangle  :  P \times Q \mapsto \mathbb{R}$ is the bilinear form
of the duality between $P$ and $ Q$. Then the linear mapping  $J :  E \ni  ( q ,  p )  \mapsto  ( p ,  q )  \in  E^\ast$ is an isomorphism, and
we identify  $h \in  E$ with  $Jh \in E^\ast$. In particular, for each
$h \in E$, the symbol $\widehat{h}$ denotes the pseudodifferential
operator on $\mathcal{L}_2(Q, \mu)$ whose Weyl symbol\footnote{The definition of a pseudodifferential operator $\widehat{F}$
on $\mathcal{L}_2(Q, \mu)$ with symbol  F can be found in [5].} is the function $Jh \in E^\ast$. 
By $\mu$ we denote the $P$-cylindrical (Gaussian) measure on $\mathcal{L}_2(Q, \mu)$ whose Fourier transform  $\Phi_\mu: P \rightarrow \mathbb{R}$ is determined by $ \Phi_\mu(p): = \exp \left(-
\frac{1}{2}\left\langle p, B_\mu p \right\rangle \right)$, where
$B_\mu: P \rightarrow Q$ is a continuous linear mapping such that
$ \langle  p , B ╡ p \rangle  > 0$, for  $p\neq  0 $. By  $\nu$ we denote a  $Q$-cylindrical
measure on  $P$ whose Fourier transform  $\Phi_\nu: Q \rightarrow \mathbb{R} $ is
defined by  $\Phi_\nu(q) : = \exp \left( -\frac{1}{2} \left\langle B^*_\mu q, q
 \right\rangle \right)$.
In what follows,
we assume that all LCSs are Hilbert, although the
main results can be extended to the general case. We
identify the space $Q$ with $ Q^\ast$ and $P$ with $ P^\ast$, so that
$B_\mu^\ast =  B_\mu$ and $ B_\mu  > 0$; note also that  $\mu$ and $\nu$ are  $\sigma$-additive if the operator  $B_\mu$ is nuclear.
The Weyl operator  $\mathscr{W}(h)$ generated by an element
$h \in  E is$ defined by  $\mathscr{W}(h): = e^{- {\rm
i}\widehat{h}}$. The Weyl function
corresponding to a density operator $ T$ is the function
$\mathcal{W}_T(h):E \mapsto \mathbb{R}$  defined by $\mathcal{W}_T(h): =
\operatorname{tr}\left(T \mathscr{W}(h)\right)$  (see [4]);
it does not depend on  $\mu$.

{\bf Definition 1} (\cite{KoSm2011}). 
The Wigner measure corresponding to a density operator  $T $ is an $ E^\ast$-cylindrical
measure $W_T$ on $ E$ determined by the relation
\begin{eqnarray*} \int_{Q \times P} e^{
{\rm i}(\left\langle p_1, q_2 \right\rangle + \left\langle p_2,
q_1 \right\rangle)}W_T(dq_1, dp_1) = \mathcal{W}_T(h)(q_2,p_2). 
\end{eqnarray*}

In other words, $W_T$ is the (inverse) Fourier transform of the function $ \mathcal{W}_T(h)$.
Therefore we have $ W_T(dq, dp) = \int_Q \int_P \mathcal{W}_T(h)(q_2, p_2) F_{E
\times E}(dq_2, dp_2, dq, dp), $ where $F_{E\times E}$  is the Hamiltonian Feynman
psuedomeasure $E \times E$.

The Feynman pseudomeasure $F_\mathcal{K}$ эon a Hilbert
space is a distribution (in the sense of the theory of
SobolevЦSchwartz generalized functions) on $\mathcal{K}$, on a Hilbert
space is a distribution (in the sense of the theory of
SobolevЦSchwartz generalized functions) on
$\mathcal{K}$. It is convenient to specify
such a functional $F_\mathcal{K}$, as well as an ordinary measure, in
terms of its Fourier transform
$\widetilde{F}_\mathcal{K}:\mathcal{K}\ni z \mapsto
F_\mathcal{K}(\varphi_z) \in \mathbb{C}$, where $\varphi_z:
\mathcal{K} \rightarrow\mathbb{C}$ is defined by
$\varphi_z(x): = e^{{\rm i}\left\langle z, x \right\rangle}$.

If $\mathcal{K} = E = Q \times P$ and
$\widetilde{F}_\mathcal{K}(q,p) = e^{{\rm i}\left\langle q, p
\right\rangle}$, then $F_ \mathcal{K}$ is
said to be a Hamiltonian Feynman pseudomeasure; it
is convenient for defining the Fourier transform that
on functions given on infinite dimensional spaces and
maps them to measures. Actually, the Hilbert space
structure is not important here; a Feynman
pseudomeasure, as well as a Gaussian measure, can be
defined on any LCS; in particular, a Hamiltonian
Feynman pseudomeasure can be defined on any symplectic LCS (additional information is contained
in [3, 9, 11]).

{\bf Proposition 1} (see \cite{KoSm2011}). If $G$ is the Weyl symbol of a
pseudodifferential operator on 
$\mathcal{L}_2(Q,
\mu)$, then $ \int_P \int_Q G(q, p) W_T(dq, dp) = \mathrm{tr}
\left(T\widehat{G}\right). $

This proposition can also be used as a definition
(cf. [4, Definition 3], where it is, however, assumed
that  $\dim Q = \dim P < \infty$ and, for this reason, only
Wigner function, rather than measures, are considered).

{\bf Definition 2.} The density $\Phi_T$ of the Wigner measure $W_T$
with respect to $\eta$ on $Q\times P$ (if this density
exists) is called the Wigner  $\eta$-function (if $\dim Q =
\dim P < \infty$ and $\eta$ is a Lebesgue measure on $Q\times P$, Єю 
then the Wigner $\eta$-function is the classical Wigner function).

In what follows, we assume that $\eta=\mu\otimes\nu$, but refer
to the Wigner
$\mu\otimes\nu$-function simply as the Wigner
function.

{\bf Corollary 1.}  If the assumptions of Proposition  1  hold,
then
\begin{eqnarray*}
\int_P \int_Q G(q,p) \Phi_T(q,p) \mu\otimes \nu (dq, dp) =
\operatorname{tr}(T\widehat{G}).
\end{eqnarray*}

{\bf Proposition 2.} The following relation holds:
\begin{eqnarray*} \Phi_T(q,p) &:=&
e^{\frac{1}{2} \left( \left\langle p_1, B_\mu^{-1}p_1
\right\rangle +
\left\langle q_1, B_\mu^{-1}q_1 \right\rangle \right)} 
\int_{Q\times P} e^{-{\rm i} \left( \left\langle p_1, q_2
\right\rangle +\left\langle p_2, q_1 \right\rangle \right)}\, \mathcal{W}_T(h)(q_2, p_2) \\
&& \times 
 e^{\frac{1}{2} \left( \left\langle p_2,
B_\mu^{-1}p_2 \right\rangle + \left\langle q_2, B_\mu^{-1}q_2
\right\rangle \right)} (\mu \otimes \nu)(dq_2, dp_2) .
\end{eqnarray*}

The function  $ (q,p) \mapsto e^{-\frac{1}{2} \left( \left\langle p,
B_\mu^{-1}p \right\rangle + \left\langle q, B_\mu^{-1}q
\right\rangle \right)} $  is the
generalized density of the Gaussian measure $\mu \otimes \nu$ (see \cite{MoSm2014} 
and the references therein). The relations
given above and those similar to them can be obtained
by using the following heuristic rule. First, we write the
corresponding formulas for the case where  $\dim
Q < \infty$, replacing Gaussian measures by their densities with
respect to Lebesgue (=Liouville) measures on the
spaces $Q$ ш $Q\times P$; in turn, these formulas are
obtained by using the standard isomorphisms between
the spaces of functions square integrable with respect
to the Lebesgue measure and the spaces of functions
square integrable with respect to the Gaussian measures. After this, we pass to the infinite dimensional
case, for which purpose we replace the Gaussian density with respect to the Lebesgue measures 
by generalized densities. It should be borne in mind that the generalized densities of Gaussian 
measures are defined
only up to multiplication by a positive number, so that
the above method for extending formulas to the infinite dimensional case applies only to formulas 
invariant with respect to the multiplication of Gaussian densities by positive numbers.
The following propositions can be regarded as definitions of Wigner measures and functions similar to
those given in [4].

{\bf Proposition 3.} For any density operator $T$ on
$\mathcal{L}_2(Q, \mu)$ and for $\varphi \in \mathcal{L}_2(Q, \mu)$,
the following relations hold:
\begin{eqnarray*} (T\varphi)(q) &=& e^{\frac{1}{4}
\left\langle B_\mu^{-1}q, q \right\rangle} \int_P \int_Q e^{-{\rm
i}\left\langle p, q_1-q \right\rangle} \varphi(q_1)
e^{-\frac{1}{4} \left\langle B_\mu^{-1}q_1, q_1 \right\rangle}\,  W_T
\left(\frac{dq_1 + q}{2}, dp\right); \\
(T\varphi)(q) &=&
e^{\frac{1}{4} \left\langle B_\mu^{-1}q, q \right\rangle} \int_P
\int_Q e^{-{\rm i}\left\langle p, q_1-q \right\rangle}
\varphi(q_1) e^{\frac{1}{4} \left\langle B_\mu^{-1}q_1, q_1
\right\rangle} \, \Phi_T \left(\frac{q_1 + q}{2}, p\right) \\
&& \times
e^{\frac{1}{2} \left\langle B_\mu^{-1}p, p \right\rangle} (\mu
\otimes \nu)(dq, dp). 
\end{eqnarray*}

The notation in the first formula means that the
mapping $q \mapsto
W_T \left(\frac{dq_1 + q}{2}, dp\right)$ is a function, while the
mapping $(dq_1, dp) \mapsto W_T \left(\frac{dq_1 + q}{2},
dp\right)$ is a measure.
The function $q \mapsto e^{-\frac{1}{2}
\left\langle B_\mu^{-1}q, q \right\rangle}$ is a generalized density of
the Gaussian measure $\mu$, and $p \mapsto
e^{-\frac{1}{2} \left\langle B_\mu^{-1}p, p \right\rangle}$ is a generalized density of the measure $\nu$.

Let$\rho_T^1$ be the integral kernel of a density
operator $T$ on $\mathcal{L}_2(Q, \mu)$, defined by 
\begin{eqnarray*} (T\varphi)
(q) = e^{\frac{1}{4} \left\langle B_\mu^{-1}q, q \right\rangle}
\int_Q e^{\frac{1}{4} \left\langle B_\mu^{-1} q_1, q_1
\right\rangle} \varphi(q_1)  \rho_T^1(q, q_1) \mu(dq_1) .
\end{eqnarray*}

{\bf Proposition 4.}  For any $\varphi \in
\mathcal{L}_2(Q,\mu)$, the following relation holds
\begin{eqnarray*} \Phi_T(q,p) =
e^{\frac{1}{2}\left(\left\langle B_ \mu q, q \right\rangle +
\left\langle B_ \mu p,p \right\rangle \right)} \int_Q \rho_T^1
\left(q - \frac{1}{2}r, q+ \frac{1}{2}r \right) e^{{\rm i}
\left\langle r,  p\right\rangle} e^{\frac{1}{2}\left\langle
B_\mu^{-1}r, r \right\rangle} \mu(dr).
\end{eqnarray*}

Let $\rho_T^2$ be the integral kernel of a density
operator $T $ on $\mathcal{L}_2(Q, \mu)$, defined by $ (T\varphi)(q) =
e^{\frac{1}{4}\left\langle B_\mu^{-1}q,q \right\rangle} \int_Q
\varphi(q_1) e^{-\frac{1}{4}\left\langle B_\mu^{-1}q_1,q_1
\right\rangle} \rho^2_T(q, dq_1) $. As such, $\rho_T^2$ is a
function of a point with respect to the first argument
and a measure with respect to the second argument

It follows from Proposition 1 that $ \rho_T^2(q, dq_1) = \int_P e^{-
{\rm i}\left\langle p, q_1-q \right\rangle} W_T \left(\frac{dq_1 +
q}{2}, dp\right) $. Setting  $s-r=q$, $s+r=q_1$ and using the chamnnge of variable
formula, we obtain $ \rho_T^2(s-r, ds+r) =
\int_P e^{- {\rm i}\left\langle p, 2r \right\rangle} W_T(ds, dp)
$, or
\begin{eqnarray*} \rho_T^2\left(q - \frac{r}{2}, dq + \frac{r}{2} \right) =
\int_Pe^{- \left\langle p, r \right\rangle} W_T(dq, dp) ; 
\end{eqnarray*}
which means that the \lq\lq measure\rq\rq  $dp \mapsto W_T( dq, dp)$ is the
inverse Fourier transform of the function $r \mapsto
\rho_T^2\left(q-\frac{r}{2}, dq + \frac{r}{2} \right)$. This implies the following proposition.

{\bf Proposition 5.} Let $F_{E}$ be a Hamiltonian pseudomeasure Feynman on $E:=Q \times P$. Then
\begin{eqnarray*}
W_T(dq, dp) = \int_Q \rho^2_T\left( q - \frac{r}{2}, dq + \frac{r}{2}\right)
F_E(dr, dp);
\end{eqnarray*}
Here, the integration with respect to the \lq\lq measure\rq\rq $dq\mapsto W_T(dq, dp)$
requires using the so-called Kolmogorov integral
\footnote{The Kolmogorov integral is the trace on the tensor product of
the space of functions on  $Q$ and the space of measures on  $Q$; $\rho_T^2$
is an element of this space (the initial definition, in which neither tensor product nor trace are mentioned, can be found in \cite{Loeve1977}).
}.

{\bf 2. EVOLUTION OF WIGNER FUNCTIONS
AND MEASURES.} 
We use the assumptions and notation of the preceding section. Suppose that, for each $t \in\mathbb{R}$, $W_T(t)$is the
Wigner measure describing the state of a quantum system at time $t$ (thus in
this section $W_T(\cdot )$ denotes a
function of a real argument whose values are Wigner
measures, while in the preceding section, the symbol $W_T(\cdot )$ denotes a Wigner measure). Then 
$W_T(\cdot )$ satisfies the equation  \cite{KoSm2011}:
\begin{equation}
\label{wigner_equ}
\dot{W}_T(t) = 2 \sin \left(\frac{1}{2} \mathcal{L}^\ast_\mathcal{H}(W_T(t))\right),
\end{equation}
where $a \in \mathbb{R}$, $\sin\left(a
\mathcal{L}^\ast_\mathcal{H}\right)$  is the linear operator
acting on the space $\mathcal{H}$ of $E^*$-cylindrical measures on $E$,
and conjugate to the operator $\sin\left(a
\mathcal{L}_\mathcal{H} \right)$, acting on the
function space on $E$, according to
\begin{eqnarray*}
\sin\left(a \mathcal{L}_\mathcal{H} \right): =
\sum_{n=1}^\infty \frac{a^{2n-1}}{(2n-1)!} \mathcal{L}_\mathcal{H}^{(2n-1)}.
\end{eqnarray*}

Here, $\mathcal{L}_\mathcal{H}^{(n)}$ is defined as follows: for each function
$\Psi: E \rightarrow \mathbb{R}$ and each  $n \in
\mathbb{N}$,
$
\mathcal{L}_\mathcal{H}^{(n)} \Psi(x) := \{\Psi,\mathcal{H}\}^{(n)}(x),
\quad x \in E,
$
where
$
\{\Psi,\mathcal{H}\}^{(n)}(x): =
\Psi^{(n)}(x) I^{\otimes n} \mathcal{H}^{(n)}(x),
$
$\Psi^{(n)}$, $\mathcal{H}^{(n)}$ denote the  $n$th derivatives of the functions$\Psi$ in $\mathcal{H}$, 
respectively, and $I^{\otimes n}$ is the $n-$th tensor power of
the operator $I$, determining the symplectic structure on
the phase space $E$
(\cite{KoSm2011}).

Relation \eqref{wigner_equ} implies an equation describing the
evolution of the Wigner $\mu$-function. To obtain it, is suffices to recall that, for any function $\Phi: E \rightarrow
\mathbb{R}$, the $n$-derivative of the product $\Phi^n\mu$ can be calculated by
the Leibniz rule and that the derivatives of the Gaussian measure $\mu$ can be calculated as follows.
If $h, h_1, h_2, \ldots \in B_\mu^\frac{1}{2} Q$, then $ \mu'h = -
\left\langle B_\mu^{-1}h,\cdot \right\rangle \mu; \mu'' h_1 h_2 =
- \left\langle B_\mu^{-1}h_1, h_2 \right\rangle\mu + \left\langle
B_\mu^{-1}h_1, \cdot \right\rangle \left\langle B_\mu^{-1}h_2,
\cdot  \right\rangle\mu,  \text{etc.} $

These relations are versions of the Wick formulas.
For each $k\in B_\mu^\frac{1}{2} Q$ the symbol $\left\langle B
_\mu^{-1}k, \cdot \right\rangle$  denotes a function defined $\mu$-almost everywhere
on $Q$, with the following properties (see \cite{Smolyanov1966})

\begin{enumerate}
\item[(i)] its domain is a measurable vector subspace of $ Q$ of full measure;
\item[(ii)] this function is linear on its domain;
\item[(iii)] if $x \in B_\mu^\frac{1}{2}Q$, then
$\left\langle B_\mu^{-1}k, x \right\rangle = \left\langle
B_\mu^{-\frac{1}{2}} k, B_\mu^{-\frac{1}{2}} x\right\rangle$
 (such
a function exists and any two functions with properties
(i)Ц(iii) coincide $\mu$-almost everywhere (see \cite{Smolyanov1966})).
\end{enumerate}

For every $a>0$, юthe operator $\sin \left(a L_
\mathcal{H}^*\right)$, cting on
functions given on $E$,
is defined by $ \sin \left(a L_\mathcal{H}^*
\right)\varphi (\mu\otimes \nu) : = \left(\sin a
\mathcal{L}_\mathcal{H}^\ast \right)(\varphi \mu\otimes\nu) $.
Suppose also that, for each$t \in\mathbb{R}$,  $\Phi_T(t)$is the Wigner $\mu$-function describing the
state of a quantum system at time  $t$.

{\bf Theorem 1.} The mapping $\Phi_T(\cdot )$,  taking values in the
set of Wigner $\mu$- functions satisfies the equation
\begin{eqnarray*}
\dot{\Phi}_T(t) = 2 \sin \left(\frac{1}{2} L_\mathcal{H}^\ast
\left(\Phi_T(t) \right)\right).
\end{eqnarray*}

{\bf 3. REDUCED EVOLUTION
OF WIGNER MEASURES}  Let $\rho_T^1$ ш
$\rho_T^2$
be the aforementioned integral
kernels of a density operator $T$  of a quantum system
being the quantum version of a classical Hamiltonian
system with phase space 
$E_1 \times E_2$, where $E_1 = Q_1 \times P_1$, and $E_2 = Q_2\times P_2$.
Then, for the integral kernels of the
reduced density operator
$T_1,$ acting on $\mathcal{L}_2(Q_i, \mu_i)$, $i=1,2$ (here and in what follows, we use the natural generalizations of the above notation and assumptions), we
have 
\begin{eqnarray*}
\rho^1_{T_1}
(q_1^1, q_2^1) &=& \int_{Q_2} \rho_T^1(q_1^1,q_2^1, q^1, q^2)
e^{\frac{1}{2}\left\langle B_{\mu_1 \otimes \mu_2}(q^1, q^2),
(q^1, q^2) \right\rangle}\, \mu_2(dq_2),\\ 
\rho^2_{T_1} (q^1,
dq_2^1)&=& \int_{Q_2} \rho_{T}^2(q^1, dq^1_2, q^2, dq^2) ;
\end{eqnarray*}
the last integral is again a Kolmogorov integral.
Therefore, Propositions 4 and 5 imply the following
theorem.

{\bf Theorem 2}  Let $W_T$ and $\Phi_T$
be the Wigner measure
and function of the quantum system with Hilbert space $\mathcal{L}_2(Q_1 \times Q_2, \mu_1 \otimes
\mu_2)$. Then the Wigner measure $W_{T_1}$ and
the Wigner function $\Phi_{T_1}$ of its subsystem with Hilbert space
$\mathcal{L}_2(Q_1, \mu_1)$ are determined by the relations 
\begin{eqnarray*}
W_{T_1}(dq_1, dp_1) &=& \int_{Q_2 \times P_2} W_T(dq_1, dp_1, dq_2,
dp_2), \Phi_T(q_1, p_1)\\
&=&  e^{\frac{1}{2}\left(\left\langle
B_{\mu_1}^{-1}q_1, q_1 \right\rangle + \left\langle
B_{\mu_1}^{-1}p_1, p_1 \right\rangle \right)}\int_{Q_2
\times P_2} e^{\frac{1}{2}\left(\left\langle B_{\mu_2}^{-1}q_2,
q_2 \right\rangle + \left\langle B_{\mu_2}^{-1}p_2, p_2
\right\rangle \right)} \\ 
&& \times \Phi_T(q_1, p_1, q_2, p_2) (\mu_2 \otimes
\nu_2)(dq_2, dp_2).
\end{eqnarray*}

Now, consider the models mentioned in the introduction. Throughout the rest of the paper, given any
Hilbert space  $\mathcal{T}$ we denote by $\mathcal{L}^s( \mathcal{T})$ the set of all
self-adjoint operators on $\mathcal{T}$.

Therefore, let $\mathscr{P}$, $\mathscr{P}_1$, $\mathscr{P}_2$,
$\mathscr{C}$, $\mathscr{C}_1$, $\mathscr{C}_2$ be Hilbert
spaces. We assume that $\mathscr{P}$ - is the Hilbert space of a
quantum control system, which we call a quantum
plant (QF), and let $\mathscr{C}$ be the Hilbert space of another
quantum control system, which we call a quantum
controller (QC); suppose that $\mathscr{P}_j$, and
$\mathscr{C}_j$, $j=1,2$ are
the Hilbert space of parts of the QP and QC, respectively. Let $\mathscr{H}: = \mathscr{P} \otimes
\mathscr{C}$ be the Hilbert space of the
united quantum system. Consider $\widehat{\mathcal{H}}_\mathscr{P} \in
\mathcal{L}^s(\mathscr{P})$, $\widehat{\mathcal{H}}_\mathscr{C}
\in \mathcal{L}^s(\mathscr{C})$,
$\widehat{\mathscr{K}}_{\mathscr{P}_1 \otimes \mathscr{C}_1} \in
\mathcal{L}^s(\mathscr{P}_1 \otimes \mathscr{C}_1)$,
$\widehat{\mathscr{K}}_{\mathscr{P}_2 \otimes \mathscr{C}_2} \in
\mathcal{L}^s(\mathscr{P}_e \otimes \mathscr{C}_2)$. We set
$\widehat{\mathcal{H}}_{\rm feedback}: =
\widehat{\mathcal{H}}_\mathscr{P} \otimes \mathbb{I}_\mathscr{C}
+ \mathbb{I}_\mathscr{P} \otimes
\widehat{\mathcal{H}}_\mathscr{C} +
\widehat{\mathscr{K}}_{\mathscr{P}_1 \otimes \mathscr{C}_1}
\otimes \mathbb{I}_{\mathscr{P}_2 \otimes \mathscr{C}_2} +
\mathbb{I}_{\mathscr{P}_1 \otimes \mathscr{C}_1} \otimes
\widehat{\mathscr{K}}_{\mathscr{P}_2 \otimes \mathscr{C}_2} \in
\mathcal{L}^s(\mathscr{H})$, уфх $\mathbb{I}_\mathscr{P} \in
\mathcal{L}^s(\mathscr{P})$, $\mathbb{I}_\mathscr{C} \in
\mathcal{L}^s(\mathscr{C})$, $\mathbb{I}_{\mathscr{P}_1 \otimes
\mathscr{C}_1} \in \mathcal{L}^s(\mathscr{P}_1 \otimes
\mathscr{C}_1)$, $\mathbb{I}_{\mathscr{P}_2 \otimes
\mathscr{C}_2} \in \mathcal{L}^s(\mathscr{P}_2 \otimes
\mathscr{C}_2)$, are the
identity operators on the corresponding spaces. The
first term in the expression for $\widehat{\mathcal{H}}_{\rm
feedback}$ describes the
evolution of an isolated QP, the second term describes
the evolution of the isolated QC, and the last two terms
describe the (coherent) quantum feedback. It is worth
mentioning that the definition of $\widehat{\mathscr{H}}_{\rm feedback}$ is symmetric with respect to QP, QC, and the feedback.

The more general Hamiltonian $\widehat{\mathcal{H}}: =
\widehat{\mathcal{H}}_{\mathscr{P}} \otimes
\mathbb{I}_{\mathscr{C}} + \mathbb{I}_{\mathscr{P}}
\otimes\widehat{\mathcal{H}}_{\mathscr{C}} +
\widehat{\mathscr{K}}$, where $\widehat{\mathscr{K}} \in
\mathcal{L}^s(\mathcal{P} \otimes\mathscr{C})$ ((see [6]),
may describe coherent quantum control both with and
without feedback. In particular, if $\widehat{\mathscr{K}} =
\widehat{\mathscr{K}}_{\mathscr{P}_1 \otimes \mathscr{C}_1}
\otimes \mathbb{I}_{\mathscr{P}_2 \otimes \mathscr{C}_2} +
\mathbb{I}_{\mathscr{P}_1 \otimes \mathscr{C}_1} \otimes
\widehat{\mathscr{K}}_{\mathscr{P}_2 \otimes \mathscr{C}_2}$), then we obtain the
previous model. On the other hand, if
$\widehat{\mathscr{K}}: = \widehat{\mathscr{K}}_{1} \otimes
\mathbb{I}_{\mathscr{P}_2 \otimes \mathscr{C}_2}$,  then we obtain a model of (coherent) quantum control without feedback.

If the QP and QC are obtained by quantizing
Hamiltonian systems, then we can assume that, in the
natural notation, $\mathscr{P}_{j}
=\mathcal{L}_2(Q_{\mathscr{P}_j},\mu_j)$, $\mathscr{C}_{j}
=\mathcal{L}_2(Q_{\mathscr{C}_j},\nu_j)$, $\mathscr{P} =
\mathcal{L}_2(Q_{\mathscr{P}_1} \times Q_{\mathscr{P}_2}, \mu_1
\otimes \mu_2)$, $\mathscr{C} = \mathcal{L}_2(Q_{\mathscr{C}_1}
\times Q_{\mathscr{C}_2}, \nu_1 \otimes \nu_2)$, $j=1,2.$  In this case, the Wigner function
and measure of the union of the QP and the QC are
defined on the space $Q_{\mathscr{P}_1} \times Q_{\mathscr{P}_2}\times
Q_{\mathscr{C}_1} \times Q_{\mathscr{C}_2}$, and their
evolution is described by the equations of the second
section. To obtain the dynamics of the Wigner function and measure of the QP (which are defined on
$Q_{\mathscr{P}_1} \times Q_{\mathscr{P}_2})$, we must apply Theorem 2.

{\bf remark 1.} Obtaining the dynamics of a quantum
control system (QP) requires finding the Hamiltonians  
$\mathscr{K}_{1}$ ш $\mathscr{K}_{2}$ (or $\mathscr{K}$)  (in appropriate classes of
Hamiltonians). This problem is similar to the simpler
problem of choosing a time dependent Hamilton
function $\mathscr{K}_{1}(\cdot)$ on $Q_{\mathscr{P}}$, to which the required dynamics
in
$\mathcal{L}_2(Q_{\mathscr{P}}, \mu)$, corresponds under the assumption
$\widehat{\mathscr{H}} = \widehat{\mathcal{H}}_{1} +
\widehat{\mathscr{K}_{1}(t)}$, where$\widehat{\mathcal{H}}_1 \in
\mathcal{L}^s(\mathscr{P})$, $\widehat{\mathscr{K}(t)}\in
\mathcal{L}^s(\mathscr{P})$. Although this model is not a special case of any of the
models described above, we expect that it can be
obtained as the limit of an appropriate sequence of
these models.

{\bf remark 2.} We can extend our model, assuming that
the QP interacts also with one more quantum system
perturbing the dynamics of the control system. Of
course, we can also assume that the source of perturbations is a part of the QP.

{\bf remark 3.}  The approach presented in the first two
sections applies directly to quantum systems obtained
by applying SchrЎdinger quantization to classical
Hamiltonian systems. To consider more general cases,
such as spin system, we must extend our approach by
methods of superanalysis. We expect that all our results
can be generalized to this case.

{\bf remark 4.} Feedback for classical Hamiltonian systems can be defined in a similar way.

{\bf remark 5.} The internal dynamics of the QP and
QC in our quantum model with (coherent) feedback
can be described in more detail. In particular, it can be
assumed that
$ \widehat{\mathscr{H}}=
\left(\widehat{\mathscr{H}}_{\mathscr{P}_1} \otimes
\mathbb{I}_{\mathscr{P}_2}  +
\mathbb{I}_{\mathscr{P}_1}\otimes
\widehat{\mathscr{H}}_{\mathscr{P}_2} \right)
\otimes\mathbb{I}_\mathscr{C}+ \mathbb{I}_\mathscr{P} \otimes
\left( \widehat{\mathscr{H}}_{\mathscr{C}_1} \otimes
\mathbb{I}_{\mathscr{C}_2} + \mathbb{I}_{\mathscr{C}_1}
\otimes \widehat{\mathscr{H}}_{\mathscr{C}_2} \right)
+\widehat{\mathscr{K}}_{\mathscr{P}_1 \otimes \mathscr{P}_2}
\otimes \mathbb{I}_{\mathscr{C}_1 \otimes \mathscr{C}_2}
+\mathbb{I}_{\mathscr{P}_1 \otimes \mathscr{P}_2} \otimes
\widehat{\mathscr{K}}_{\mathscr{C}_1 \otimes \mathscr{C}_2}
 \quad  + \widehat{\mathscr{K}}_{\mathscr{P}_1 \otimes
\mathscr{C}_1} \otimes \mathbb{I}_{\mathscr{P}_2 \otimes
\mathscr{C}_2} + \widehat{\mathscr{K}}_{\mathscr{P}_2 \otimes
\mathscr{C}_2} \otimes \mathbb{I}_{\mathscr{P}_1 \otimes
\mathscr{C}_1}. $

In the above relation, the parts of the Hamiltonian
describing the QP and the QC and the interaction
between them are again symmetric.

{\bf  ACKNOWLEDGMENTS}
This work was begun while the authors were visiting
the Isaac Newton Institute for Mathematical Sciences
(Cambridge, UK) within the framework of the Program УQuantum Control Engineering: Mathematical
Principles and Applications,Ф of which J. Gough was
one of the organizers. Gough acknowledges the support of the British EPSRC, grant EP/L006111/1.
T.S. Ratiu acknowledges the support of Swiss National
Scientific Foundation, NCCR SwissMAP and grant
no. 200021Ц140238. O.G. Smolyanov acknowledges
the support of the Russian Foundation for Basic
Research, project no. 14-01-00516.

{\footnotesize

}

\end{document}